\documentstyle[multicol,prl,aps,epsf,psfig,epsfig]{revtex}

\begin{document}

\title
{Phase Transitions in an Aging Network.}
\author
{Kamalika Basu Hajra and  Parongama Sen}
\address
{
Department of Physics, University of Calcutta,
    92 Acharya Prafulla Chandra Road, Kolkata 700009, India. \\
}
\maketitle
\begin{abstract}

We consider a growing network in which an incoming  
node gets attached to the $i^{th}$ existing  node with
the probability  $\Pi_i \propto {k_i}^{\beta}\tau_i^{\alpha}$, 
where $k_{i}$ is the degree of the $i^{th}$ node and $\tau_i$ 
its present age. The  phase diagram in the ${{\alpha}-{\beta}}$ 
plane is obtained. The network shows scale-free behaviour, i.e., 
the degree distribution $P(k) \sim k^{-\gamma}$ with $\gamma =3$
only along a line in this plane. Small world property, on the 
other hand, exists over a large region in the phase diagram.

\end{abstract}

\noindent Preprint no. CU-Physics-04/2004.

\begin{multicols}{2}

     Complex web-like structures describe a wide variety of systems of
high technological and intellectual importance. The statistical properties
of many such networks have been studied recently with much interests.
Such networks, with complex topology are common in nature and examples
include the world wide web, the Internet structure, social networks, 
communication networks, neural networks to name a few \cite{BArev,DMbook,Watts_book}.

 Some of the common  features which are of importance in these networks of 
diverse nature are:\\

  (i) Diameter of the network: This is defined as the maximal shortest path over all vertex pairs. The networks in which the diameter
$(\cal D)$  varies as the logarithm of the number of nodes $(N)$, i.e, 
${\cal {D}} \propto ln(N)$, are said to have the {\it small world }(SW)
 property. On the other
hand if $\cal D$ varies as a power of $N$ 
we get what can be termed as {\it regular networks}. 0ne can also study $D$, the average 
shortest distance between pairs of nodes. In general $D$ and  $\cal D$ have the same scaling behaviour.\\

  (ii) Clustering coefficient: A common property of many real networks 
is the tendency  to form clusters or triangles,  
quantified by the {\it clustering coefficient}. 
This is in contrast to random networks (Erd\"os and R\'enyi\cite{ER}) where small world property
is present but the clustering tendency is absent.\\

  (iii) Degree Distribution: The node 
 degree  distribution function $P(k)$ 
 gives
the probability that a randomly selected node has exactly $k$ edges,
In a random network this is a Poisson distribution. In many real world 
networks $P(k)$ 
  shows a power law decay and
  such networks are termed  
{\it Scale Free Networks}. \\
  
In order to emulate the different features of real networks 
several models have been proposed. While the Watts-Strogatz  \cite{WS}
network provides an appropriate model for the small world network (i.e., 
small diameter and finite clustering coefficient),
 scale free properties of a network can be reproduced by models proposed 
later by Barab\'asi and Albert (BA) \cite{BA} and independently by 
Hubermann and Adamic\cite{HA}.

	In the BA model, a network is grown from a few nodes and new nodes are added one by one. At a time $t$, the incoming node  
is connected randomly to the  $i^{th}$ existing node
 with the attachment probability ${\Pi}_{i}(t)$  given by 
\begin{equation}
{\Pi}_{i}(t)\sim {k_{i}(t)},
\end{equation}
where $k_{i}$  is the degree of the $i^{th}$ node.
The degree distribution in the BA model shows the scale-free behaviour 
\begin{equation}
P(k)\sim{k^{-\gamma}},
\end{equation}
with ${\gamma}=3$.

  Following its introduction, several modifications in the BA model have been 
studied. A few of them are worth mentioning here in the context of the present paper.  
  Non-linear dependence of the attachment probability 
on $k$, in the model designed by Krapivsky {\it et al} (KRL)\cite{KRL},  
shows that the scale-free property exists only for 
the linear dependence. This nonlinear model has been studied in much 
detail very recently in \cite{Onody}. On the other hand, BA model on an  Euclidean network
\cite{Manna_Sen,senmanna}  has
also  been considered in which the attachment probability 
has been modified as follows:

\begin{equation}
\Pi_{i}(t)\sim {k}_{i}(t)^{\beta}l^{\delta},
\end{equation}
where $l$ is the Euclidean length between the new and old  nodes.
 A phase diagram in the
${\beta}-{\delta}$ plane was obtained along with other 
  interesting features.\\

Another important modification in the BA model has been made by
incorporating time dependence in the network \cite{DM,Zhu}.
In real life networks, a time factor
may also modulate the  attachment probability. In most of the real world 
networks, aging of the nodes usually takes place, e.g.,  one rarely cites 
old papers, or in social networks, people of the same age are more likely to be linked. Dorogovtsev and Mendes (DM) \cite{DM} studied the case when the
 connection probability of the new site with an old one is not only
proportional to the degree $k$ but also to the power of 
its present age, ${\tau}$, such that 
\begin{equation}
{\Pi}_{i}(t)\sim {k}_{i}(t){\tau }_i^{\alpha}
\end{equation}\\
 and they showed both numerically and analytically that the
scale free nature disappears when $\alpha < -1 $. Here $\alpha$ governs
 the dependence of the attachment probability on the "age difference" of two
 nodes, i.e, for negative values of $\alpha$, a new node will tend to attach
 itself to the younger nodes. Therefore for the extreme case
 $\alpha \to -\infty$, a new node will attach itself to its immediate 
predecessor while for the case $\alpha \to \infty$, the oldest
 and a few very old nodes will 
 get more edges. The time dependence presents a competing effect when
 $\alpha < 0$ but for $\alpha > 0$, the older nodes get even more 
rich, similar to the rich gets richer effect.  

Encouraged by the  existence of the various phase transitions observed in the
modified BA models, 
 we have    further generalised the time dependent
BA network.
Here we generate a network
such that the attachment probability is given by 
\begin{equation}
{\Pi}_i(t)\sim {k}_i(t)^{\beta}{\tau_i^{\alpha}}.
\end{equation}
We expect here that $\beta \ne 1$ will change the behaviour of the DM model
as in \cite {KRL}. The competing effect of $\alpha$ is able to destroy the scale free nature of the DM model ($\beta=1$). The effect of a positive $\beta (>1)$ and negative 
$\alpha$ could re-instate the scale free behaviour as in \cite{senmanna} 
and  
 it is also possible to obtain a phase diagram in the $\alpha-\beta$ plane. 
Formally eq (5) is analogous to eq (3). However here the nodes are chronological, i.e, the age of the initial node is $t$ at time $t$, that of the second node $t-1$ and so on. In the Euclidean network on the other hand
 the coordinates  of the nodes are uncorrelated.
Moreover, the dimensionality plays an important factor in it.

The known limits of this model are therefore\\
1. $\beta = 1, \alpha = 0$ - BA network\\
2. $\beta$ any value, $\alpha = 0$ - KRL  network\\
3. $\beta = 1, \alpha $ any value  - DM network\\
When $\alpha $ and $\beta $ are both zero, we get a random growing network which shows an exponential decay of $P(k)$ \cite {BArand}.

The network is generated in the usual manner where we start with a single
 node and at every time step   the new  node gets attached to one of the 
existing nodes with an attachment probability given by equation (5). 
 
 We have considered nodes with a single incoming link such that there are no loops
and the clustering coefficient is zero.  Thus  we focus our attention on the degree distribution and the 
average distance to study the small world and 
scale free behaviour.

 From  eq (5) we predict that for any $\beta$  as ${\alpha}\to+{\infty}$  
 a {\it gel} formation is expected  when majority of the nodes tend 
to get attached to   a 
particular node. On the other hand when ${\alpha}\to-{\infty}$ we expect  a 
 {\it regular chain} formation (in the time space) when each node gets attached to its immediate predecessor.
The average shortest distance $(D)$ 
 for both ${\alpha}\to+{\infty}$  and ${\alpha}\to-{\infty}$ are easy to
calculate.  
When $\alpha\to-\infty$, $D$ is given by
\[
 D=\frac{{\Sigma_{1}^{N}}(k(k-1)+(N-k)(N-k+1))}{2N(N-1)}.
\]
\[
 = (N+1)/3.
\]
On the other hand, for large values of
$\alpha$,
$D$  has a finite value $\sim O(1)$.
Hence it is natural to 
expect a transition from a small world 
behaviour to a regular behaviour as $\alpha$ is varied. In fact for $\beta=0$, 
one can locate approximately  the transition point using some simple arguments. 

\begin{center}
\begin{figure}
\includegraphics[clip,width= 4cm,angle=270]{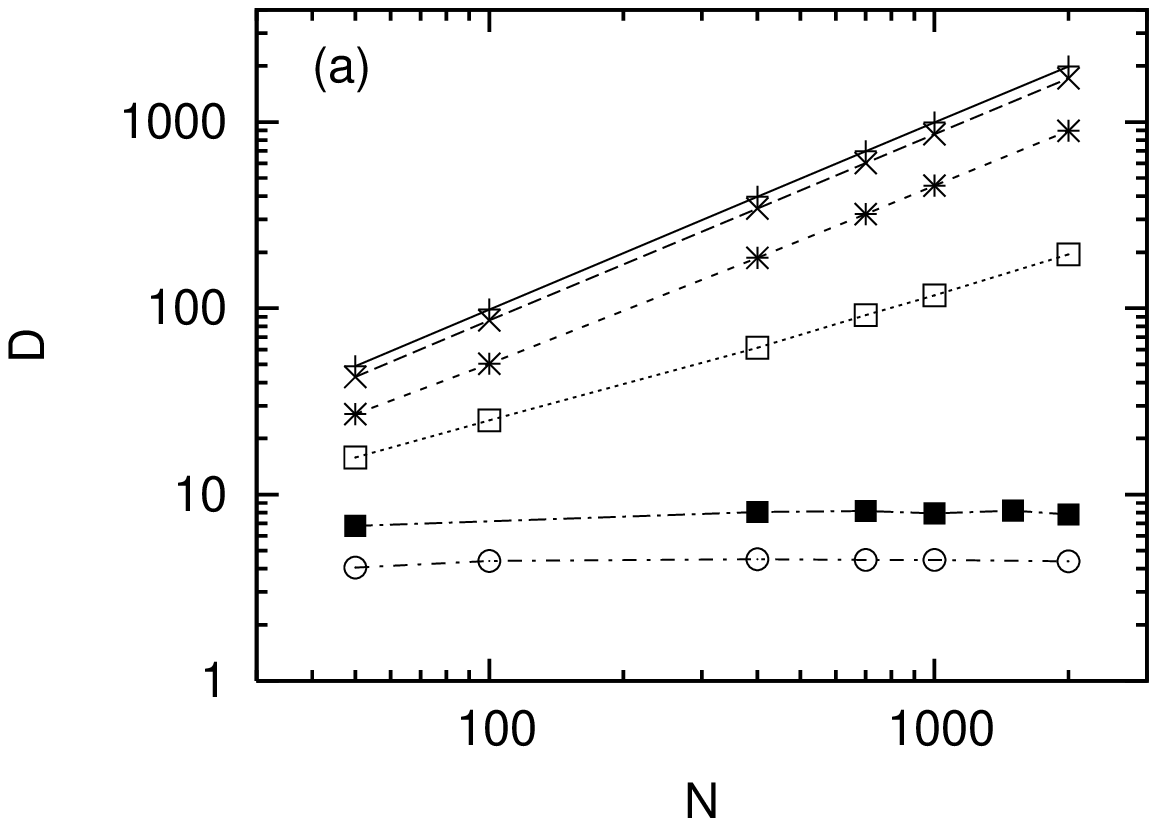}
\includegraphics[clip,width= 4cm,angle=270]{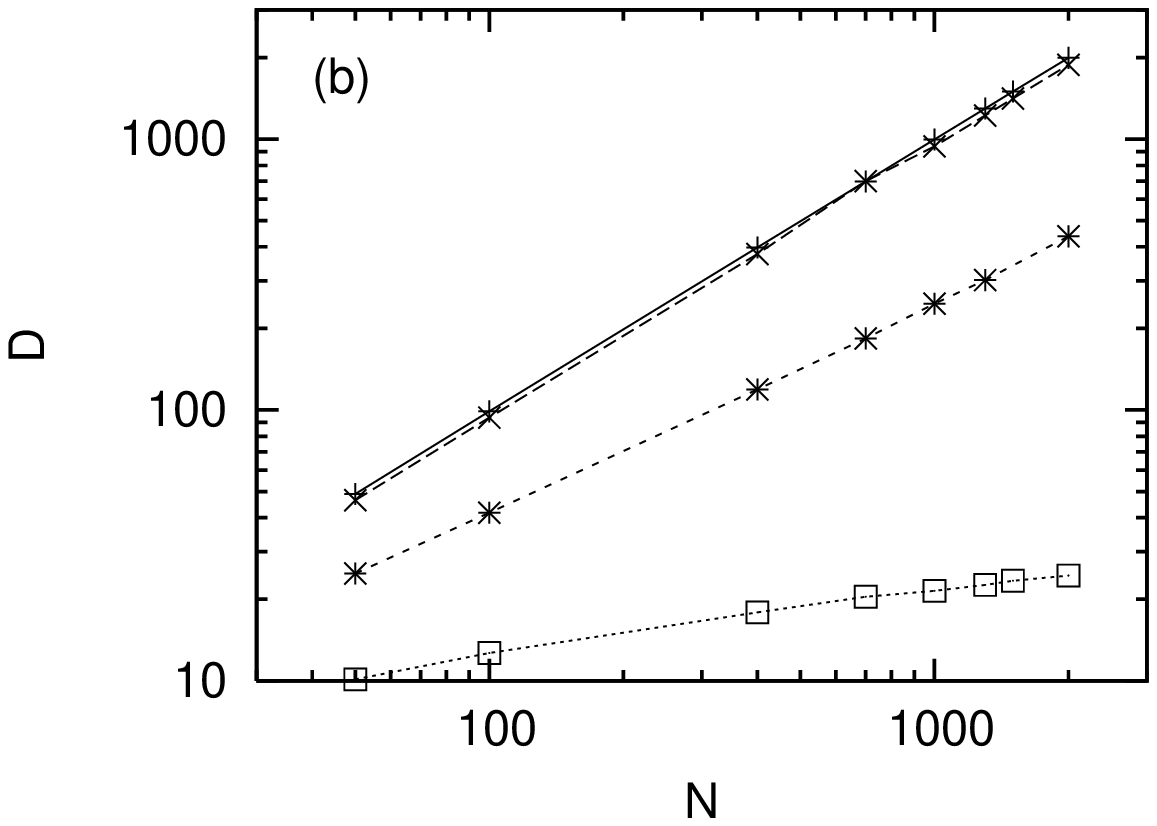}
\caption{The variation of the average shortest distance $D$ with $N$ 
for various values of $\alpha$ at (a)$\beta=2$ and (b)$\beta=0.5$. In (a) the
 exponent $\lambda$ changes value from $1$ to $0$ sharply as
we go from top to bottom of the figure, corresponding to 
$\alpha=-10,-5,-3,-2,0$. In (b) $\lambda$ changes from $1$ to a relatively small value as $\alpha=-10,-5,-2,0$ from top to bottom.}
\end{figure}
\end{center}  

In analogy 
with \cite{bond}, one can define here an ``age difference factor'' $\Delta\tau_{ij}=|t_i-t_j|$ if the $i^{th}$ node of age $t_i$ and $j^{th}$ node of age $t_j$ are connected. If the network has been evolved till a 
time $t(\geq2)$, then for the incoming node we can write
\begin{equation}
{\langle\Delta\tau\rangle}_t=\frac{\int_1^t\tau^{\alpha+1}{d\tau}}{\int_1^t\tau^\alpha{d\tau}}.
\end{equation}  
 For the small world property, the behaviour of
 $\langle \Delta\tau \rangle$ for large $t$ should be studied. From
 eq(6), for large $t$, $\Delta\tau\sim O(1)$ for $\alpha<-2$
 and therefore 
there can be no small world behaviour for $\alpha<-2$ for large networks. 
On the other hand for $\alpha>-1$, $\langle\Delta\tau\rangle\sim O(t)$, from
which one can expect SW property for $\alpha>-1$. We in fact find 
a small world to regular network transition  
at $\alpha=-1$ numerically.

   The simulations have been made using a maximum of $2000$ nodes for 
studying small world properties and $4000$ nodes for determining degree 
distribution, using a maximum of $1000$ configurations.  
   In the analysis of the small world
characteristics, we  calculate  $D$ for the networks for different values of 
${\beta}$  and ${\alpha}$.  
 The $D$ versus $N$ curve   is
generally  of the form
$D\sim N^{\lambda}$,
where the exponent $\lambda$ depends on $\alpha$ (see Fig. 1).

In order to locate the transition to the small world (where $\lambda$ 
is either zero or has a very small value)  we  note the variation of  
$\lambda$ with  ${\alpha}$
for different values of ${\beta}$. We  observe that for
 all values of ${\beta}$, there is a sharp fall in $\lambda$ from  unity  
 to a very small value 
 indicating a transition from regular to small world behaviour.
  The transition point shifts to a more negative value
of ${\alpha}$ as we move from smaller to larger  values of $\beta$.
Typical  $\lambda$ vs. $\alpha$ plots are shown in Fig. 2.

\begin{center}
\begin{figure}
\includegraphics[clip,width= 4cm,angle=270]{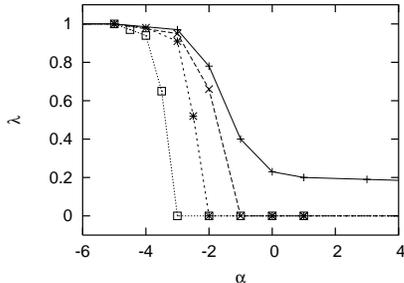}
\caption{The variation of the exponent $\lambda$
with ${\alpha}$ for different values of ${\beta}$ (from right to left of the figure, $\beta=0,2,3,4$). There is a sharp transition
in the value of ${\alpha}$ from $+1$ to $0$ as network behaviour changes from regular to small world.}
\end{figure}
\end{center}

 Next we study the degree distribution $P(k)$ for the network for several values
 of $\alpha$ and $\beta$. For the regular chain limit ($\alpha \to -\infty$),
most of the vertices have degree $2$, while for the gel phase ($\alpha \to
 +\infty$), there will be a very high maximal degree and many leaves (i.e, 
nodes with degree=1).
 Thus the different phases can be identified from the behaviour of $P(k)$.
 First let us discuss the known case for $\beta=1$.
We find that $P(k)$ has an exponential decay at $\alpha = -1$ as in \cite{DM}
and has scale-free (SF) behaviour for $\alpha > -0.5$. The latter value does not
agree with \cite{DM} and  the possible reasons of discrepancy are  discussed
later.
For other values of $\alpha$, $P(k)$ has a 
stretched exponential (SE) behaviour, i.e.,
\begin{equation}
P(k)\sim \exp(-ax^b),
\end{equation}
where $b$ depends on $\alpha$.   
 Allowing $\beta$ to assume values greater
than unity, we find that SF behaviour exists only for a specific value of 
$\alpha=\alpha^*$, e.g. at $\beta=3$ we obtain
 this behaviour at $\alpha=\alpha^*=2.5$ (Fig. 3).

\begin{center}
\begin{figure}
\includegraphics[clip,width= 4cm,angle=270]{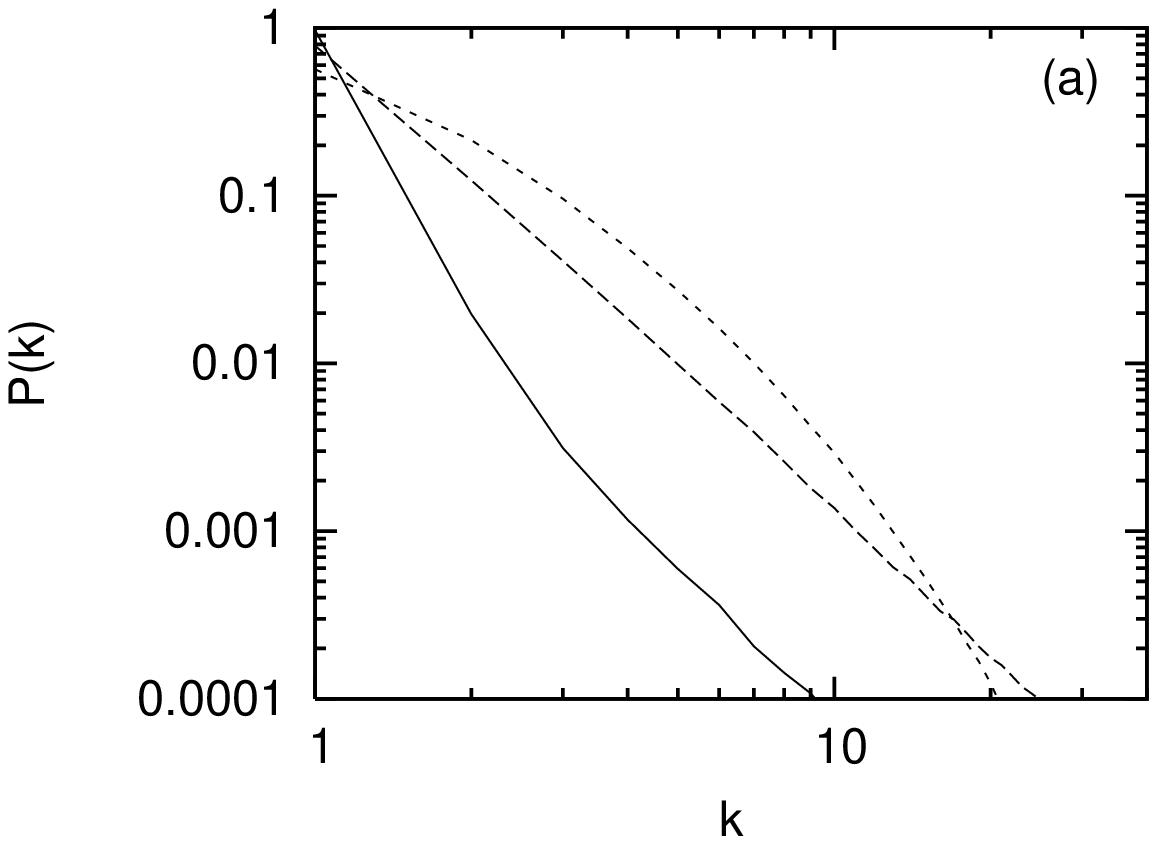}
\includegraphics[clip,width= 4cm,angle=270]{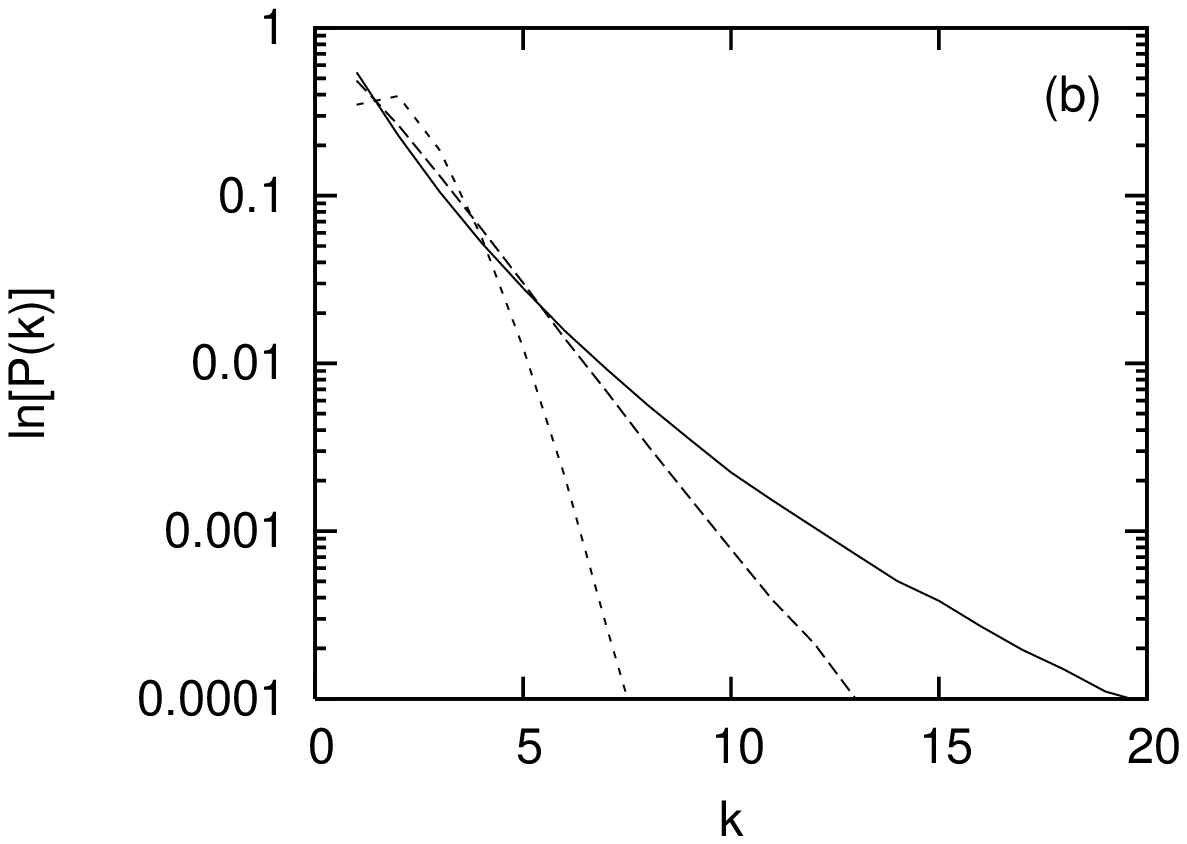}
\caption{(a) The $P(k)$ vs $k$ plot in log-log scale for  $\beta=3$. Here, at
 $\alpha=\alpha^*=2.5$ (dashed line), there is scale free behaviour, while for other 
values (e.g. $\alpha=-2.8,-2.3$) this behaviour is lost.
(b)P(k) vs k plot in log-linear scale for $\beta=0.8$. Here exponential behaviour is observed at $\alpha=-0.9$ (dashed line), while for
 other values (e.g. $\alpha=-0.5,-2.0$) stretched exponential nature is observed.}
\end{figure}
\end{center}

For $\alpha>\alpha^*$, we get a gel-like behaviour, while for $\alpha<\alpha^*$, we again get a stretched exponential  behaviour.
The scale-free behaviour for $\beta \geq 1$ always occurs with 
$\gamma =3$ as in the BA network.
  For $\beta<1$, SF behaviour is not observed for any value of $\alpha$. Here 
$P(k)$ shows a SE behaviour as in eq (7).

   In fig.4, we have shown  the phase diagram in the 
$\alpha-\beta$ plane. We have plotted the phase boundary between the 
SW and the regular network regions, the curve along which scale-free behaviour 
exists  and  the line along which  $b=1$. 
  The $b=1$ line  is 
not a phase boundary, but it has the interesting property that it has the
behaviour of a random growing network albeit with non-zero values of
$\alpha$ and $\beta$. For negative values of $\alpha$, 
aging  can be regarded as a competitive phenomenon to preferential 
attachment to the extent that one recovers the random behaviour 
even for high values of $\beta$ along this line. 

A small world network is not necessarily scale-free but  a scale-free
network is usually  a small world barring some exceptional or artificial
 cases (e.g,
if one considers $N$ number of BA networks in a series, it is a scale-free
but not a small world network).
To the right of the scale-free line and above it, the network
shows a gel-like behaviour.
Both  the scale-free and gel regions do have the small world 
property, as expected, 
but there are finite regions in the phase diagram where the degree distribution is not a power law but of 
 different types (e.g., exponential or stretched exponential)
with small world behaviour.

\begin{center}
\begin{figure}
\includegraphics[clip,width=6cm,angle=0 ]{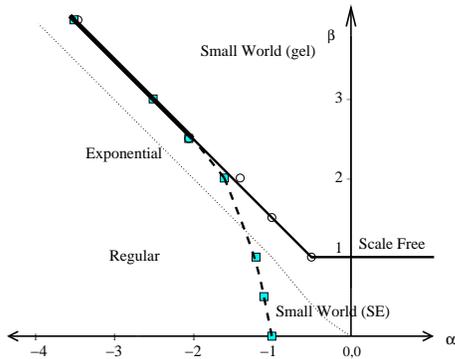}
\caption{The phase diagram for the given network in the 
${\alpha}$-${\beta}$ plane. The  small world (SW) regions with gel-like as
well as stretched exponential (SE) behaviour, the regular
chain region and the scale free region are indicated. The network is scale free along the thinner solid line while 
 the broken line is the phase boundary for SW-regular transition and these two lines merge along the thicker solid line. The dotted line is the one along which $b=1$, i.e, where the network is random in nature. (All lines are guides to the eye.)}
\end{figure}
\end{center}

   In summary, we have generalised the BA network to include time-dependent 
or aging effects in the attachment probability (eq. (5)) 
such that both the
time  dependence and the degree dependence
can be parametrically tuned. A phase diagram is obtained in the $\alpha-\beta$
 plane, where $\alpha,\beta$ are the parameters governing the two 
factors respectively. We claim that this is the most generalised network
where both time dependence and degree dependence are incorporated in the 
preferential attachment. There is a quantitative disagreement in the transition
point at $\beta=1$ as compared to \cite{DM} which may be because
of the finite sizes considered here. The time and effort required to locate
phase transition points are considerable and a finite size analysis
has not been attempted therefore.
Other   results known earlier, e.g.,  
 gel
formation beyond $\beta>1$  for  $\alpha=0$, exponential decay of $P(k)$
for both $\alpha, \beta =0$ etc.
   have been recovered in our simulations.
Similar to the Euclidean network \cite{Manna_Sen,senmanna}, the scale free behaviour 
is found to exist along a single line here.  
In fact, as regards the scale-free boundary, the present phase diagram 
is very much similar to that obtained in \cite{senmanna}.
However, here the network belongs to the BA universality class 
($\gamma=3$) along the entire line. 
Moreover,  one can  compare the present results with the
one dimensional Euclidean network only for which the phase diagram 
is available.   
 A phase boundary for small world to regular transition has also been obtained.
The network may have small world behaviour even when the degree
distribution is exponential or stretched exponential. Along the $\alpha=0$ line, the network retains the small world behaviour, consistent with the results of 
\cite{Onody}, where it was found that $D$ assumed a finite value ($\sim lnN$) 
for networks of different sizes for all values of $\beta$.

 It is worth mentioning here that the limiting forms of the degree distribution,
 at extreme values of $\alpha$, are delta functions in nature, but we have 
restricted our analysis to finite values of $\alpha$. Also, we find that the 
phase diagram shows varied features for values of $\alpha<0$ for which 
the model corresponds to realistic networks like citation, 
collaboration or social networks.\\

 Acknowledgments: We thank S. S. Manna for useful comments. KBH is grateful to CSIR (India) F.NO.9/28(609)/2003-EMR-I for financial support.
PS acknowledges DST grant no.  SP/S2/M-11/99.

Email: kamalikabasu2000@yahoo.com, parongama@vsnl.net

\end{multicols}


\begin{thebibliography}{}
\bibitem{BArev} R. Albert and A. -L. Barab\'asi, Rev. Mod. Phys. {\bf 74}, 47 (2002).
\bibitem{DMbook} S. N. Dorogovtsev and J. F. F. Mendes, {\it Evolution of Networks}, Oxford University press (2003).
\bibitem{Watts_book} D. J. Watts, {\it Small Worlds: The Dynamics of Networks between Order and Randomness}, Princeton University Press, Princeton, New Jersey (1999).
\bibitem{ER} P. Erd\"os and A. R\'enyi, {Publ. Math. Debrecen {\bf{6}}, 290 (1959)}; {Publ. Math. Inst. Hung. Acad. Sci. {\bf{5}}, 17 (1960).}
\bibitem{WS} D. J. Watts and S. H. Strogatz, Nature {\bf {393}}, 440 (1998).
\bibitem{BA} R. Albert and A. -L. Barab\'asi, {Phys. Rev. Lett. {\bf{85}}, 5234 (2000).}
\bibitem{HA} B. A. Huberman and L. A. Adamic, Nature (London) {\bf {401}} ,130 (1999).
\bibitem{KRL} P. L. Krapivsky, S. Redner and F. Leyvraz , Phys. Rev. Lett.
 {\bf {85}}, 4629 (2000).
\bibitem{Onody} R. N. Onody and P. A. de Castro, Physica A {\bf 336}, 491 (2004).
\bibitem{Manna_Sen} S. S. Manna and P. Sen, Phys. Rev. E {\bf{66}}, 066114
(2002). 
\bibitem{senmanna} P. Sen and S. S. Manna, Phys. Rev. E {\bf 68}, 026104 (2003).
\bibitem{DM} S. N. Dorogovtsev and J. F. F. Mendes, Phys. Rev. E {\bf{62}}, 1842 (2000); {\bf{63}} 056125 (2001).
\bibitem{Zhu} H. Zhu, X. Wang and J-Y. Zhu, Phys. Rev. E {\bf 68}, 056121 (2003). 
\bibitem{BArand} A. L. Barab\'asi, R. Albert and H. Jeong, Physica A {\bf 272}, 
173 (1999).
\bibitem{bond}P. Sen, K.Bannerjee and T. Biswas, Phys. Rev. E {\bf 66},
037102 (2002).
\end{thebibliography}
\end{document}